\documentclass[11pt,a4paper]{article}

\usepackage[T1]{fontenc}
\usepackage[utf8]{inputenc}
\usepackage{tipa}
\usepackage{amsmath,amssymb,amsfonts}
\usepackage{graphicx}
\usepackage{booktabs}
\usepackage{hyperref}
\usepackage{multirow}
\usepackage{cite}
\usepackage{algorithm}
\usepackage{algorithmic}
\usepackage[margin=1in]{geometry}
\usepackage{times}

\title{Tokenizations for Austronesian Language Models: study on languages in Indonesia Archipelago}

\author{%
  Andhika Bernard Lumbantobing\thanks{Bandung Fe Institute \& Adjunct Science Fellow in InaAI, \texttt{nad@compsoc.bandungfe.net}}
  \and
  Hokky Situngkir\thanks{AI Research Center IT Del \& Bandung Fe Institute, \texttt{hokky.situngkir@del.ac.id}}
}

\date{Version: 2026-01-27}

\begin{document}

\maketitle

 \begin{abstract}
Tokenization constitutes a fundamental stage in Large Language Model (LLM) processing; however, subword-based tokenization methods optimized on English-dominant corpora may produce token fragmentation misaligned with the linguistic structures of Austronesian languages. This study aimed to develop a syllable-based tokenization framework adopting principles from traditional Indonesian scripts (aksara) for regional languages of Indonesia. A syllabic segmentation procedure was constructed based on the logic of abugida writing systems and implemented with a vocabulary of 2,843 tokens extracted from the Indonesian dictionary (KBBI). Evaluation was conducted on the NusaX dataset comprising 1,000 parallel translation samples across 10 regional languages, Indonesian, and English. Analysis employed Token per Character (TPC) ratio and sequence alignment using the Smith-Waterman algorithm. Results demonstrated that syllable-based tokenization yielded consistent TPC values ($\sim$0.4) across all regional languages, whereas GPT-2 exhibited an inverse pattern with the lowest TPC for English. Syllable-based tokenization consistently produced higher token sequence similarity scores, with an average increase of approximately 21\% (slope $m \approx 1.21$) compared to GPT-2. These findings confirm that the syllable-based approach more effectively preserves phonological and morphological patterns across related Austronesian languages, offering a linguistically principled foundation for multilingual LLM development.


 \end{abstract}

\section{Introduction}
One of the critical stages in the processing of modern text-based \emph{Large Language Models} (LLMs) is tokenization, the process of decomposing human-readable text into discrete units called tokens. The operational domain of LLMs is constrained to these units. Input text is transformed into a sequence of tokens, which are subsequently mapped to numerical representations. These representations serve as the internal computational substrate that enables the model to perform algebraic operations on linguistic phenomena. During the inference or text generation phase, these numerical representations are processed through neural network layers with trained weights \cite{vaswani2017}. Language generation in LLMs is, in principle, a process of forming token combinations, where each token is selected based on contextual information contained in the preceding token combinations, thereby enabling the formation of novel linguistic objects \cite{friedman2023}. The tokenization strategy and the resulting tokens thus determine and direct the types of linguistic regularities that a language model can learn and internalize \cite{mielke2021}.

In practice, tokenization cannot be regarded as a neutral or entirely language-agnostic preprocessing step \cite{ali2023}. Modern LLMs predominantly employ subword-based tokenization schemes such as \emph{BPE} \cite{sennrich2016}, \emph{WordPiece} \cite{schuster2012}, and \emph{Unigram} \cite{kudo2018}. These methods are driven by statistical efficiency optimization objectives that are highly sensitive to the characteristics, distributions, and regularities found in their training corpora \cite{bostrom2020}, which directly determine the vocabulary inventory or subwords deemed important through training, including morphological patterns, phonotactics, and orthographic conventions. Training tokenization with these methods on corpora disproportionately dominated by a small subset of particular languages, such as English, introduces structural biases that favor certain linguistic forms \cite{rust2021, petrov2023, velayuthan2025}, while their application to languages with fundamentally different linguistic typologies may produce token fragments that are misaligned with the language structure and obscure meaningful linguistic boundaries and patterns.

The study on the development of an Indonesian Language LLM (TOBA-LM) \cite{situngkir2026} demonstrated that tokenization strategies designed based on linguistic principles can enhance language naturalness in downstream text generation performance. Following this approach, we position tokenization as a linguistically motivated design choice. This study explores opportunities for extending the application of syllabic segmentation-based approaches to regional languages in Indonesia that share similar characteristics within the Austronesian language family. The distinctive characteristics of Austronesian languages motivate the design of a cross-linguistic tokenization framework that can be applied consistently at the language family level.

In its implementation, we adopt the principles of traditional writing systems as the computational foundation for designing syllable-based tokenization schemes. Traditional scripts in Indonesia, which belong to the abugida (\emph{alphasyllabary}) writing system, inherently represent syllabic units as their basic elements. Historically, this system constituted the dominant form of orthography in literacy practices across the archipelago and can be viewed as a form of language encoding aligned with natural speech boundaries, emerging from the mechanics of speech production and the cognitive processes of its speakers \cite{friedman2023}.

The initial section of this paper discusses the linguistic characteristics of languages within the Austronesian family and the principal tenets of traditional Indonesian-ethnic based writing systems that underpin the construction of the proposed syllable-based tokenization. The subsequent section elaborates on the computational procedures for constructing the tokenization scheme along with the technical adjustments required for its integration into modern LLM architectures. To evaluate the proposed approach based on linguistic alignment criteria, this study conducts an empirical comparison with the GPT-2 subword tokenization scheme as a statistically-driven tokenization approach without explicit linguistic assumptions.

\section{Austronesian Languages and Traditional Writing Systems}

The Austronesian language family constitutes one of the largest and most geographically dispersed language families in the world, encompassing more than 1,200 languages spanning from Madagascar to Easter Island, and from Taiwan to New Zealand \cite{blust2013}. The structural similarities among Austronesian languages reflect their common origin, namely Proto-Austronesian (PAn), an ancestral language estimated to have been spoken thousands of years ago in the Taiwan region. Linguistic reconstruction indicates that the basic lexical units in PAn are canonical roots that are predominantly disyllabic. This phonotactic pattern is generally reconstructed as CV(C)CVC, where C represents a consonant and V a vowel \cite{blust2013}. This pattern remains evident in modern Austronesian languages, including Indonesian, which exhibits a relatively simple syllable structure with a maximum CVC boundary, as well as a strong tendency toward disyllabic words \cite{lapoliwa1981}.

In addition to these phonological characteristics, Austronesian languages are marked by rich and agglutinative morphological systems. In agglutinative languages, word formation is accomplished productively through the sequential addition of affixes to base forms, including prefixes, suffixes, infixes, circumfixes, and reduplication processes \cite{iacobini2006}. Functionally, these affix sequences serve as markers of complex grammatical relations, including voice marking, valency changes such as causatives, and applicative constructions. Although the Austronesian lexicon is generally built upon disyllabic roots, the application of agglutinative rules to the root inventory constitutes the primary factor in the formation of complex derived words, which may consist of three, four, or even more syllables. Indonesian likewise exhibits these characteristics. A quantitative study of Indonesian morphology \cite{denistia2022} demonstrates high productivity in affixation, reduplication, compounding, and cliticization. This agglutinative character is also found in regional languages of the Indonesian archipelago, such as Javanese \cite{poedjosoedarmo1979}, Sundanese \cite{coolsma1985}, and Batak \cite{nababan1981}.

The combination of predominantly disyllabic roots and agglutinative morphology that ``stacks'' affixes renders the syllable a stable and fundamental organizational unit in Austronesian languages. This uniformity of characteristics found across various Austronesian languages motivates the design of a cross-linguistic tokenization framework centered on syllables that can be applied to \emph{large language model} (LLM) implementations at the language family level. This framework also addresses two functional limitations inherent in phoneme-based and word-based tokenization. Tokenization schemes based solely on phonemes fail to capture the prosodic rhythm and natural structure of Austronesian languages, while whole-word tokenization would encounter an explosion of morphological variation that triggers high \emph{out-of-vocabulary} (OOV) ratios, particularly in scenarios with constrained vocabulary budgets.

In implementing the syllable-based tokenization framework, we adopt the processing logic from traditional writing systems that developed in the Indonesian archipelago, namely the Indonesian-ethnic based scripts. According to linguistic classification, these traditional writing systems are categorized as \emph{abugida} \cite{daniels1996, daniels2001} or \emph{alphasyllabary} \cite{bright1996, bright1999}, wherein the basic linear unit, termed ``aksara'' or ``akhsara'' is essentially a syllable. An aksara may consist of a consonant with an inherent vowel, a consonant with a vowel diacritic, an independent vowel, a consonant cluster with a vowel, or a consonant with a ``killed'' vowel \cite{bright1999}. This system originates from the Brahmi script of ancient India in the 3rd century BCE, which was used during the Ashokan Empire \cite{daniels2019}. Through its dissemination, the script underwent divergence and active adaptation to accommodate the distinctive linguistic contours of each region. The earliest form to enter the Indonesian archipelago was the Pallava script from South India, which subsequently developed into the Kawi script in Java-Bali and the Malay regions. Additionally, contact with traders from Gujarat who introduced the Nagari script contributed to the emergence of various local scripts in Sumatra (including Batak and Surat Ulu/Rejang) and in Sulawesi (Bugis-Makassar), which later gave rise to derivative scripts in Ende and Bima \cite{miller2010, daniels2019}. Beyond Indonesia, similar writing systems are also found extensively in other Southeast Asian languages, such as those of Myanmar, Thailand, Laos, Cambodia, and the Philippines \cite{court1996, kuipers1996}.

Despite exhibiting graphemic variations and orthographic realizations through diverse formal mechanisms according to regional specificities, language processing, particularly syllable segmentation, in Indonesian-ethnic script systems operates according to similar fundamental principles and demonstrates convergent functional patterns. Based on these commonalities, we construct tokenization procedures for Latin script representations of regional languages by abstracting and simulating the internal logic of the Indonesian-ethnic scripts.

\subsection{Aksara}

Unlike alphabetic systems that are based on segmental phonemes, aksara writing systems (and abugida systems in general) employ the syllable as the fundamental orthographic unit. Each base character (glyph) represents either an independent vowel or a consonant with an inherent vowel, typically /a/. Certain phonemes such as /ng/ and /ny/, which in Latin alphabet-based romanization are realized through the digraphs $\langle$ng$\rangle$ and $\langle$ny$\rangle$, are represented graphemically in traditional scripts as single units with inherent vowels ($\langle$\emph{ng}a$\rangle$ and $\langle$\emph{ny}a$\rangle$), as can be observed in the Javanese-Balinese, Batak, and Bugis scripts. Vowel sound modification on consonants with inherent vowels is accomplished through the addition of diacritical marks that modify the base symbol, while vowel deletion to form coda consonants in producing closed syllable sounds is achieved through vowel suppression using a special marker (\emph{virama}).

Inter-civilizational contact contributed to reconstructing the phonological inventory of local languages while simultaneously influencing their script systems as coded in the culture from which these languages originated \cite{situngkir2016}. Sanskrit influence, for instance, is manifested in the adoption of retroflex consonants in the Javanese script, which distinctly differentiates between dental series ($\langle$ta$\rangle$, $\langle$da$\rangle$) and retroflex series ($\langle$\emph{th}a$\rangle$, $\langle$\emph{dh}a$\rangle$). Furthermore, Arabic and Malay influences enriched the fricative sound inventory. To accommodate these borrowed phonemes, modified glyphs or companion characters (\emph{aksara rekan}) were developed. In the Javanese script, special symbols are found for sounds such as /x/ (kh), /\textesh/ (sy), /f/, /z/, and /q/. A similar phenomenon is observed in the Sundanese script, which developed seven modified characters to represent sounds such as $\langle$fa$\rangle$, $\langle$va$\rangle$, $\langle$qa$\rangle$, $\langle$xa$\rangle$, $\langle$za$\rangle$, $\langle$\emph{kh}a$\rangle$, and $\langle$\emph{sy}a$\rangle$.

\subsection{Medial Clusters}

In languages that permit consonant clusters, various script traditions have developed specialized mechanisms to represent consonant clusters (C1C2) without explicitly suppressing the inherent vowel. This mechanism is primarily applied to clusters where the second consonant (C2) is a semivowel or liquid, such as /r/, /l/, /y/, and /w/. This approach is found in classical Indian orthographic traditions as well as in several Indonesian ethnolinguistic traditions, such as Javanese-Balinese, which tend to avoid the use of \emph{virama} in word-medial positions. In Javanese orthography, consonant clusters involving the sounds /r/, /l/, /y/, and /w/ are represented through special diacritics known as \emph{sandhangan wyanjana}. The Sundanese script categorizes medial consonant markers under the class of \emph{rarangkén}, namely \emph{panyakra}, \emph{panyiku}, and \emph{pamingkal}. Meanwhile, in Balinese orthography, medial consonant markers are classified as \emph{pangangge aksara}. In certain Sumatran writing traditions, such as Batak and Rejang, which did not develop specialized symbols for medial consonants, consonant cluster handling is accomplished by explicitly suppressing the vowel on C1 through a vowel killer mark (\emph{virama}).

\begin{table}[h]
\centering
\caption{Consonant Cluster Handling in Traditional Scripts}
\label{tab:medial_consonants}
\begin{tabular}{lcccc}
\toprule
\multirow{2}{*}{Medials} & \multicolumn{3}{c}{Writing Systems} \\
\cmidrule{2-4}
& Javanese & Balinese & Sundanese \\
\midrule
Medial /r/ & Cakra & Guwung & Panyakra \\
\midrule
Medial /y/ & Pengkal & Nania & Pamingkal \\
\midrule
Medial /l/ & Pasangan La & Gantungan La & Panyiku \\
\bottomrule
\end{tabular}
\end{table}

\subsection{Coda Representation}

Fundamentally, aksara units with their inherent vowels represent open syllables. To construct closed syllables, traditional script systems have developed several mechanisms. In general, a coda—one or more consonant sounds that close a syllable—is realized by nullifying the inherent vowel on a consonant aksara through a special symbol known as \emph{virama}. In certain script traditions that avoid the use of \emph{virama} in word-medial positions, such as the Javanese and Balinese scripts, specialized written forms have been developed that function similarly to \emph{virama} in word-final positions. Table \ref{tab:killer_marks} illustrates the terminological differences for vowel suppression mechanisms found across several traditional scripts.

\begin{table}[h]
\centering
\caption{Vowel Killers (\emph{Virama}) in Traditional Scripts}
\label{tab:killer_marks}
\begin{tabular}{lccccc}
\toprule
\multirow{2}{*}{Vowel Killers} & \multicolumn{5}{c}{Writing Systems} \\
\cmidrule{2-6}
& Javanese & Balinese & Sundanese & Batak & Rejang \\
\midrule
Mid-Word Killer & Pasangan & Gantungan & Pamaéh & Pangolat & - \\
Final Killer & Pangkon & Adeg-adeg & Pamaéh & Pangolat & Muris \\
\bottomrule
\end{tabular}
\end{table}

In addition to vowel killer mechanisms, several traditional scripts have also developed specialized representations for certain codas, which are generally realized as diacritics that modify the base aksara. These diacritics are used to mark final consonants with high frequency of occurrence or prominent phonological function, such as the nasal consonant /ng/ or the liquid /r/. This enables the writing of closed syllables without complete reliance on generic vowel suppression mechanisms. In the Javanese script, these specific coda representations are known as \emph{sandhangan panyigeg wanda}, while in the Balinese script, a similar mechanism is found within the diacritic class called \emph{pangangge tengenan}. Table \ref{tab:final_consonants} presents the variety of nomenclature and functions of coda diacritics found in other traditional scripts, such as Sundanese, Batak, and Rejang.

\begin{table}[h]
\centering
\caption{Coda Diacritics in Traditional Scripts}
\label{tab:final_consonants}
\begin{tabular}{lccccc}
\toprule
\multirow{2}{*}{Phoneme} & \multicolumn{5}{c}{Writing Systems} \\
\cmidrule{2-6}
& Javanese & Balinese & Sundanese & Batak & Rejang \\
\midrule
Final /ng/ & Cecek & Cecek & Panyecek & Hamisaran & Tulang \\
Final /r/ & Layar & Surang & Panglayar & - & Junjung \\
Final /h/ & Wignyan & Bisah & Pangwisad & Hajoringan & Kah \\
\bottomrule
\end{tabular}
\end{table}

\section{Construction of Syllabic Segmentation for Austronesian Languages}

In this section, we elaborate the procedure for transforming Latin text into sequences of syllabic segments according to rules based on aksara orthography. For the Latin alphabet \(\Sigma\) consisting of a vowel set \(V = \{ a,\ i,\ u,\ ...\} \subset \Sigma\) and consonants (non-vowels) \(C = \Sigma\backslash V\), the segmentation procedure is defined as \(P:\ \Sigma^{*} \rightarrow T^{*}\), which maps an input symbol sequence \(S = (s_{1}...s_{n}),\ s_{i} \in \Sigma\) to an output segment sequence \(U = (u_{1}...u_{k})\). Each segment \(u_{i}\)\hspace{0pt} is an ordered tuple \((\omega,\ \nu,\ \kappa) \in T\), where \(T \subseteq \Sigma^{*}\) is a subset of alphabet symbol sequences derived through \(P\) and represents syllables, with a structure defined as the arrangement of onset component \(\omega\), nucleus \(\nu \in V \cup \{\varnothing\}\), and coda \(\kappa\).

In phonetic terminology, onset refers to the consonant or consonant cluster preceding the vowel, while nucleus constitutes the syllable core, which is generally a vowel. In the context of traditional scripts, onset corresponds to the base consonant aksara, while the following nucleus represents the inherent vowel that may undergo sound modification or suppression. In closed syllable formations, the nucleus is followed by a consonant cluster termed coda. Several traditional script systems encode certain codas using specialized diacritics that modify the base aksara; these special codas are formalized as the set \(H = \{ ng,\ h,\ l,\ ...\} \subset C^{*}\). Members of set \(H\) are consonant sequences that are primarily treated as ``attached'' to the preceding nucleus and do not form a new onset.

To accommodate consonant digraph forms in the Latin alphabet that are represented as single aksara in traditional orthography, we define the digraph set \(D = \{ th,\ dh,\ ...\}\  \subset C^{2}\) as consonant sequences that are treated as indivisible or as a single unit. Additionally, semivowels that potentially form medial consonant clusters and receive specialized symbolization in traditional scripts are defined by the set \(M = \{ r,\ l,\ w,\ ...\} \subset C\). Medial cluster handling follows the modification mechanism on base aksara, which is applied only if the semivowel \(s_{i} \in S,s_{i} \in M\) satisfies two conditions: the following alphabet symbol \(s_{i + 1} \in V\) is a vowel, and the preceding \(s_{i - 1}\) is a consonant or valid digraph that is neither a member of \(M\) (to avoid consecutive semivowels) nor a member of the special coda set \(H\).

The segmentation process begins with a linear scanning phase over the input symbol sequence \(S\) from left to right to extract potential phonological units. During iterations in this phase, onset candidates (\(\omega\)), whether single consonants, digraphs, or medial cluster formations, are evaluated. The vowel symbol following onset \(\omega_{j}\) is designated as nucleus \(\ \nu_{j}\), and if nucleus \(\nu_{j}\) is followed by a sequence that is a member of the special coda set \(H\), that sequence is identified as coda \(\kappa_{j}\) for the corresponding syllabic unit to form the complete unit tuple \(u_{j} = (\omega_{j},\ \nu_{j},\ \kappa_{j})\). In conditions where a consonant or digraph is not directly followed by a vowel, that sequence is designated as onset for an isolated unit without nucleus or coda \(u = (\omega,\varnothing,\varnothing)\). This unit will be handled in the subsequent phase following the vowel killer (\emph{virama}) principle in traditional scripts. After a unit is defined, scanning continues to evaluate the onset for the next syllabic unit. This process repeats until the end of sequence \(S\) is reached. Onset identification for syllabic units can be formally expressed as follows.

\begin{equation}
\omega(i) = 
\begin{cases} 
\alpha\beta, & \beta \in M \wedge S[i + |\alpha| + 1] \in V \wedge \alpha \notin (H \cup M) \\ 
\alpha, & \text{otherwise} 
\end{cases}
\end{equation}

and

\begin{equation}
\alpha = 
\begin{cases} 
S[i : i + 2], & S[i : i + 2] \in D \\ 
S[i], & \text{otherwise} 
\end{cases}
\end{equation}

where \(V\) denotes vowels, \(M\) semivowel consonants, \(D\) consonant digraphs, and \(H\) the special coda set. After the syllabic unit sequence is identified by the initial phase, the procedure proceeds to handle units without nucleus \(\nu(u) = \varnothing\). In traditional scripts, these units reflect consonant aksara that have undergone inherent vowel suppression and become codas in closed syllables. In this phase, backward iteration \((k,k - 1,...2)\) is applied to sequence \(U = (u_{1}...u_{k})\) to identify units without nucleus and integrate them as coda components of the preceding syllabic unit:

\begin{equation}
\nu(u_{i}) = \varnothing \Rightarrow \kappa(u_{i - 1}) \leftarrow \kappa(u_{i - 1}) + \omega(u_{i})
\end{equation}

where \(\nu\), \(\kappa\), and \(\omega\) respectively denote the nucleus, coda, and onset of a syllabic unit. This \emph{virama} mechanism simulation phase is then followed by a final phase that handles remaining units without nucleus. These units are generally complex consonant clusters in modern loanwords that deviate from local phonotactics and whose orthography is not yet accommodated by traditional rules. In this phase, forward iteration \((1,2,...k - 1)\) is applied to the syllabic sequence resulting from the previous phase. Each identified unit without nucleus is merged into the onset component of the following syllabic unit:

\begin{equation}
\nu(u_{i}) = \varnothing \Rightarrow \omega(u_{i + 1}) \leftarrow \omega(u_{i}) + \omega(u_{i + 1})
\end{equation}

The text segmentation procedure \(P\) elaborated above can be viewed as a composition of three processing phases. The first phase \(\phi_\text{scan}:\ \Sigma^{*} \rightarrow T^{*}\) maps alphabet sequences into pre-tokens by identifying base aksara along with their inherent vowels as potential phonological units. The output of this processing is continued by the second transformation \(\phi_\text{vir}:\ T^{*} \rightarrow T^{*}\), which reflects the principle of vowel suppression mechanism. The transformation \(\phi_\text{clus}:\ T^{*} \rightarrow T^{*}\) is then applied to handle vowel-less consonant clusters not accommodated by the preceding processing. Thus, the function \(P\) that produces syllabic segment sequences can be written as the following composition function:

\begin{equation}
P = (\phi_\text{clus} \circ \phi_\text{vir} \circ \phi_\text{scan})
\end{equation}

\section{Evaluation}

\subsection{Tokenization}

To evaluate the proposed syllabic segmentation procedure, we implemented it within a tokenization scheme that adjusts to syllable distribution in language dictionaries. In this regard, we utilized the word list from the Kamus Besar Bahasa Indonesia (KBBI, the Great Dictionary of the Indonesian Language) as the training corpus to determine the base token vocabulary. Following the previous approach in \cite{situngkir2026}, the segmentation procedure was first applied to the corpus to identify syllable units and estimate their unigram distribution. Syllables with high frequency of occurrence, which generally reflect local dependencies among character symbols in natural language text, were selected as tokens in vocabulary \(\Sigma\), along with base symbols such as Unicode characters. To handle \emph{Out-of-Vocabulary} (OOV) cases for the fraction of syllables not accommodated as tokens, we applied a \emph{fallback} mechanism that decomposes syllable units into their constituent characters, which are mapped to base symbols in \(\Sigma\). In this construction, the tokenization \(\tau_{\Sigma}:\mathcal{X}^{*} \rightarrow {\mathcal{T}_{\Sigma}}^{*}\) that maps Latin alphabet sequences to token sequences can be written as:

\begin{equation}
\tau_{\Sigma} = (f_{\Sigma} \circ P)
\end{equation}

where \(P\) is the syllabic segmentation function in Equation (5), \(\Sigma\) is the vocabulary constructed based on training, and \(f_{\Sigma}\) maps syllable segments \(u \notin \Sigma\) to single character sequences \(u_{1},...,u_{k} \in \Sigma\).

\begin{figure}[h]
\centering
\includegraphics[width=15.92cm,height=11.22cm]{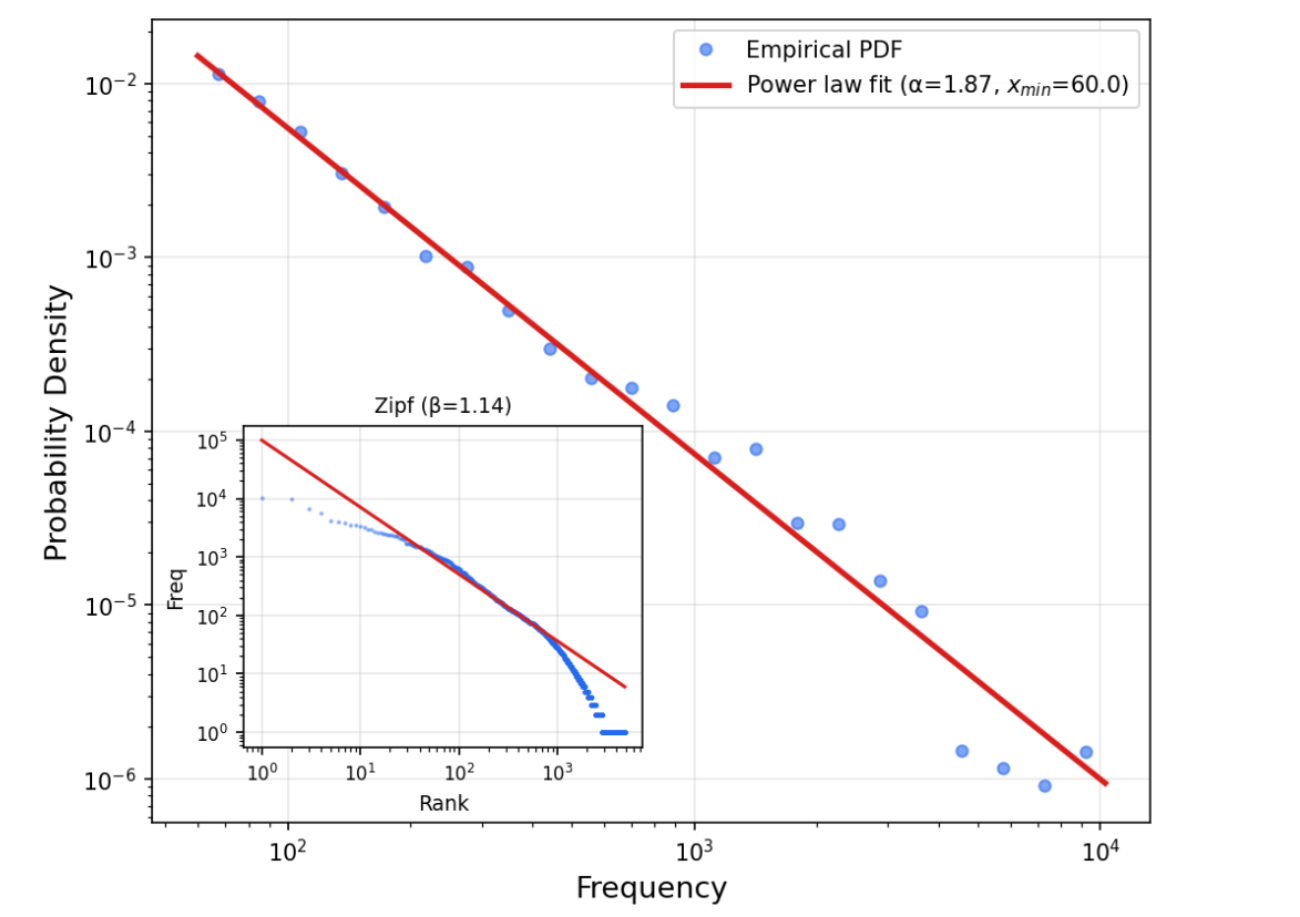}
\caption{The frequency distribution of syllable occurrences in the KBBI corpus follows a power law $p(x) \sim x^{-\alpha}$ with coefficient $\alpha \approx 1.87$. \textbf{Inset}: Syllable rank distribution with $\beta = \frac{1}{\alpha-1} \approx 1.14$.}
\label{fig:figure1}
\end{figure}

In the evaluation phase, we employed syllable-based tokenization with a vocabulary size of \(|\Sigma| = 2,843\). The composition of this vocabulary comprises syllables with the highest frequency of occurrence, monosyllabic lexemes, and base character symbols. Although the vocabulary was constructed exclusively based on the Indonesian language corpus distribution, testing was conducted on corpora involving regional languages as a cross-linguistic evaluation. As a baseline comparison, we utilized the GPT-2 \emph{tokenizer} \cite{radford2019}, which has a vocabulary size of \(50,257\). The GPT-2 \emph{tokenizer} is a \emph{subword}-based tokenization scheme constructed using the \emph{Byte Pair Encoding} algorithm \cite{gage1994} to optimize byte sequence compression on 40GB of Internet text data in the WebText training corpus.

\subsection{Data}

Evaluation was conducted using the NusaX dataset \cite{winata2023}, a multilingual parallel corpus encompassing 10 regional languages in Indonesia: Acehnese (ace), Balinese (ban), Banjarese (bjn), Toba Batak (bbc), Buginese (bug), Javanese (jav), Madurese (mad), Minangkabau (min), Ngaju (nij), and Sundanese (su), along with their Indonesian and English equivalents. This dataset was developed by translating SmSA \cite{purwarianti2019}, an Indonesian sentiment analysis dataset containing comments and reviews, to support sentiment analysis and machine translation tasks. We specifically utilized the machine translation subset (NusaX-MT) and employed all 1,000 available samples (comprising the combined 500 training samples, 100 validation samples, and 400 test samples) for analytical purposes. Table \ref{tab:lang_stats} presents statistics on the number of unique words for each language in the NusaX corpus.

\begin{table}[h]
\centering
\caption{Unique Word Count Statistics in NusaX}
\label{tab:lang_stats}
\small
\begin{tabular}{lcccccccccccc}
\toprule
\textbf{id} & \textbf{en} & \textbf{ace} & \textbf{ban} & \textbf{bbc} & \textbf{bjn} & \textbf{bug} & \textbf{jav} & \textbf{mad} & \textbf{min} & \textbf{nij} & \textbf{su} \\
\midrule
4,259 & 4,277 & 4,239 & 4,914 & 4,658 & 4,621 & 5,121 & 4,708 & 4,863 & 4,454 & 3,986 & 4,688 \\
\bottomrule
\end{tabular}
\end{table}

Analysis in the evaluation was performed on tokenization outputs. The data preparation phase was conducted by processing each sample in the corpus using both the syllable-based tokenizer and GPT-2. This phase produced token sequences as encoded representations of each sample in the respective languages.

\subsection{Methods}

\subsubsection{Segmentation Compatibility and Cross-Linguistic Syllables}

The multilingual effectiveness of syllable-based tokenization schemes can be measured based on their representational capacity when applied to target languages. Within the framework of the proposed tokenization procedure, evaluation can be conducted on the compatibility of tokens trained on the Indonesian corpus against regional languages. The presence of language-specific syllable segments or segmentation patterns misaligned with the linguistic characteristics of the target language may cause candidate segments to be unaccommodated in the tokenizer vocabulary. This condition is subsequently handled by the fallback mechanism, which decomposes such segments into single character symbols, thereby producing longer token sequences. For a symbol sequence \(S = (s_{1},...s_{k})\) as text in a target language, this can be quantified by calculating the \emph{Token per Character} (TPC) ratio as follows:

\begin{equation}
TPC(S) = \frac{|\tau_{\Sigma}(S)|}{|S|}
\end{equation}

where \(|\tau_{\Sigma}(S)|\) is the length of the encoded representation of \(S\) after tokenization by \(\tau_{\Sigma}\), and \(|S|\) is the sequence length prior to tokenization. TPC values in target languages exhibiting significant differences from the source language value (in this case, Indonesian) indicate misalignment between the tokenization scheme and the target language structure.

\subsubsection{Token Sequence Alignment}

Tokenization schemes aligned with linguistic features enable models to exploit lexical and morphological similarities across languages, as well as facilitate semantic representation alignment in \emph{joint training} strategies for multilingual LLMs within the same language family \cite{pires2019, artexte2020}. Misalignment between tokenization schemes and the linguistic structure of the languages they represent can distort relevant linguistic boundaries \cite{hofmann2021} and diminish the potential benefits of cross-linguistic learning \cite{patil2022}. Approaches that rely solely on full lexical overlap among cognate languages tend to overlook internal variation and divergence, including local phenomena such as dialectal differences. Conversely, syllable-based approaches that decompose words into sublexical units have the potential to reveal shared patterns in token sequences across languages with similar typologies.

In the context of language translation, these shared patterns may manifest as cognates and dialectal variations, and their identification requires approaches sensitive to sound shifts and affixation changes. Structurally, this is analogous to mutation, insertion, and deletion phenomena in genetic sequences. In bioinformatics, \emph{sequence alignment} was developed to align DNA, RNA, or protein sequences to identify regions of similarity while handling mismatches, point mutations, and gaps that may arise from evolutionary divergence \cite{mount2004, jones2004}. Significant sequence similarity following alignment can indicate common ancestry or functional similarity. To evaluate the proposed tokenization scheme, we adopt the application of \emph{sequence alignment} on tokenization outputs to measure the extent to which the tokenization scheme produces segments aligned with meaningful linguistic patterns and capable of capturing similarity in token sequence pairs across languages with similar typologies.

Specifically, we employ the local alignment method, which enables detection of significant local similarity regions between sequence pairs that are globally divergent. This approach focuses on conserved fragments while disregarding portions with low similarity, and permits measurement of linguistic similarity despite differences in token position or sequence length. We implemented this local alignment using the Smith-Waterman algorithm \cite{smith1981, gotoh1982}.

For token sequence pairs resulting from the same tokenization procedure \(\tau_{\Sigma}(S_{1}) = a_{1}a_{2}...a_{n}\) and \(\tau_{\Sigma}(S_{2}) = b_{1}b_{2}...b_{m}\), we construct a scoring matrix \(H\) of dimensions \((n + 1) \times (m + 1)\) and initialize the first column and row with zeros, \(H_{k,0} = H_{0,l} = 0\) for \(0 \leq k \leq n\) and \(0 \leq l \leq m\). This initialization allows the alignment process to commence from any position in the sequences without incurring initial penalties. The procedure then continues by populating scores for each remaining matrix element \(H\) in order, from left to right and top to bottom. This process considers results from substitution (diagonal score) as well as gap insertion (horizontal and vertical scores). The score at each element \(H_{i,j}\)\hspace{0pt} represents the maximum similarity of subsequences ending at tokens \(a_{i}\) and \(b_{j}\)\hspace{0pt}. This value is obtained through the following recursive relation:

\begin{equation}
H_{ij} = \max_{}\{ H_{i - 1,j - 1} + s(a_{i},b_{j}),\ \max_{k \geq 1}(H_{i - k,j} - w_{k}),\ \max_{l \geq 1}(H_{i,j - l} - w_{l}),\ 0\}
\end{equation}

for \(1 \leq i \leq n\) and \(1 \leq j \leq m\), where \(s(a_{i},b_{j})\) is the match score if tokens \(a_{i}\) and \(b_{j}\) are identical or the penalty score if they differ, while \(w_{k}\) and \(w_{l}\) are respectively the penalties for gaps of length \(k\) in sequence \(\tau_{\Sigma}(S_{1})\) and length \(l\) in \(\tau_{\Sigma}(S_{2})\). The use of zero as the lower bound ensures non-negative scores to disregard sequence portions with low similarity. In the evaluation conducted, we applied a scoring scheme that prioritizes exact matches while maintaining flexibility for gaps and token mismatches, using the following weights:

\begin{equation}
s(a_{i}, b_{j}) = 
\begin{cases} 
2, & a_{i} = b_{j} \\ 
-1, & a_{i} \neq b_{j}
\end{cases}
\end{equation}

\begin{equation}
w_{k} = w_{l} = - 1\
\end{equation}

The maximum value in the matrix, \(\max_{}(H)\), is identified as the highest score marking the terminal position of subsequences from \(\tau_{\Sigma}(S_{1})\) and \(\tau_{\Sigma}(S_{2})\) with the most optimal match among all possible subsequences. This score is used as a similarity measure between corresponding sentence pairs in two different languages, and is normalized against the geometric mean of the maximum potential scores of each sequence to neutralize the influence of sequence length. The similarity value between text pairs \(S_{1}\) and \(S_{2}\) is formulated as follows:

\begin{equation}
\text{sim}(S_1, S_2) = \frac{\max(H)}{2 \times \sqrt{|\tau_{\Sigma}(S_1)| |\tau_{\Sigma}(S_2)|}}
\end{equation}

where \(H\) is the scoring matrix calculated based on the weights in Equations (9) and (10).

\subsection{Results}

\begin{figure}[h]
\centering
\includegraphics[width=15.49cm,height=9.91cm]{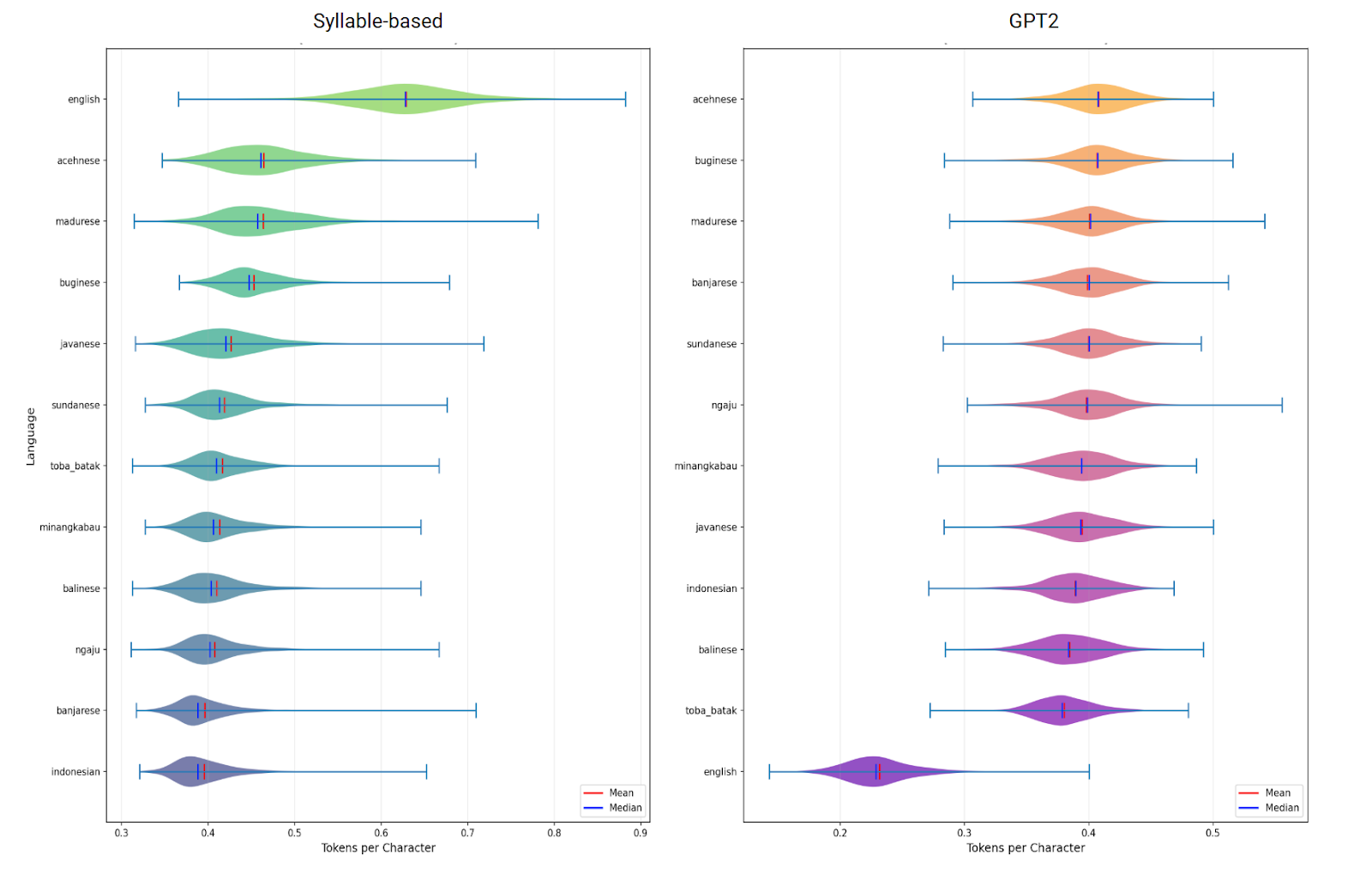}
\caption{Comparison of \textit{tokens per character} distributions between the syllable-based tokenization method (left) and GPT-2 tokenization (right) across various languages. The red and blue vertical lines represent the mean and median values of the distributions, respectively.}
\label{fig:figure2}
\end{figure}

Evaluation of \emph{Tokens per Character} (TPC) values reveals contrasting performance patterns between the syllable-based tokenization scheme and GPT-2, as illustrated in Figure \ref{fig:figure2}. Although the vocabulary for syllable-based tokenization was extracted exclusively from the Indonesian corpus, the TPC values produced across various regional languages demonstrate a high degree of uniformity, approaching the Indonesian TPC value. This indicates that the proposed aksara-based tokenization procedure can be applied consistently across languages and yields segmentation results exhibiting an ``internal compatibility'' among regional language tokens rooted in their shared Austronesian heritage. Quantitatively, the stable TPC values hovering around 0.4 confirm the characteristic of Austronesian languages dominated by CV(C) syllabic patterns with an average length of 2 to 3 phonemes per unit.

Meanwhile, the application of syllable-based tokenization to English produced TPC values that were deviant and significantly higher compared to the regional languages. This increase in TPC values is a consequence of the high frequency of fallback mechanism activation, triggered by the incompatibility of English syllabic segments with the Indonesian-based vocabulary inventory. Empirically, this highlights fundamental differences in phonotactic patterns and linguistic characteristics between analytic language families, such as English, and the languages in the region of Indonesian archipelago.

An inverse pattern was observed in TPC values resulting from GPT-2 tokenization, where English obtained the lowest TPC value while regional languages exhibited relatively higher values. Optimization on a corpus predominantly dominated by English produces a GPT-2 subword vocabulary that is longer and more efficient for this language. Meanwhile, the minimal representation of regional languages in the training data causes their linguistic units to be fragmented into shorter subwords. Although the resulting TPC values are relatively comparable to syllable-based tokenization, this efficiency is achieved through a different mechanism, namely compression optimization based on high-frequency character sequence grouping. Given that compression approaches are sensitive to training data distribution, such groupings can be arbitrary and may not represent meaning or linguistic function in regional languages.

\begin{figure}[h]
\centering
\includegraphics[width=15.92cm,height=7.62cm]{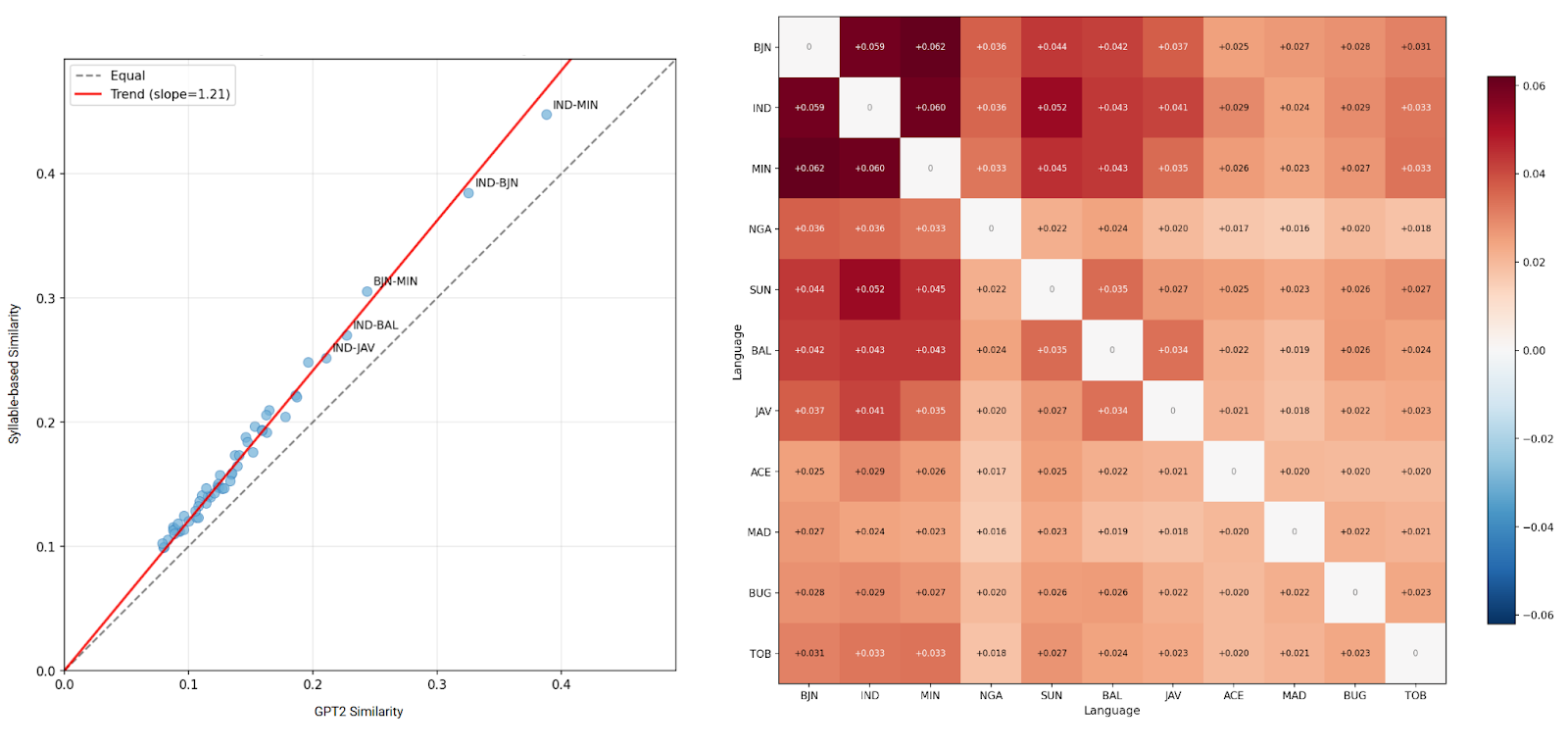}
\caption{Comparative analysis of token sequence similarity values between the syllable-based method and GPT-2. (a) Scatter plot comparing similarity across language pairs. (b) Heatmap of similarity value differences between the two tokenization schemes (Syllable-based -- GPT-2) for each language pair. Red color gradation indicates higher similarity values obtained by the syllable-based method.}
\label{fig:figure3}
\end{figure}

Furthermore, evaluation based on sequence alignment reveals differences in the degree of linguistic preservation and alignment performance between syllable-based tokenization and GPT-2 tokenization. Token sequence alignment was performed on each parallel translation sample, and cross-linguistic similarity was calculated as the mean similarity value across all samples for each tokenization scheme. Figure \ref{fig:figure3}a displays the comparison of similarity values for token sequences of language pairs produced by both tokenization approaches. Overall, evaluation of these proximity values demonstrates a consistent inter-language relational structure regardless of the tokenization scheme employed. This is evidenced by the strong linear correlation (\(r = 0.9956\)) and monotonic relationship (\(\rho = 0.9895\)) between language pair values obtained from both approaches.

Nevertheless, syllable-based tokenization consistently produces higher similarity values for all regional language pairs compared to GPT-2 tokenization. The slope trend of \(m \approx 1.21\) in Figure \ref{fig:figure3}a indicates an average increase of approximately 21\%. This finding suggests that cross-linguistically relevant linguistic patterns are better preserved through syllable tokenization, while GPT-2 tokenization fragments these patterns into more arbitrary segments. Figure \ref{fig:figure3}b demonstrates that the difference in proximity values between the two tokenization schemes is most pronounced in language subgroups that, according to traditional linguistic scholarship, share close genealogical relationships. Minangkabau and Banjarese, which belong to the Malayic language family, exhibit the highest value differences as well as stronger proximity to Indonesian. This is attributable to the fact that syllable-based tokenization was trained with an Indonesian vocabulary inventory, rendering it more sensitive to the structures of cognate languages within the Malayic family. Beyond the Malayic family, increased similarity values were also observed in the language cluster that developed in Java and its surrounding regions. Specifically, syllable-based tokenization identifies a higher degree of proximity between Balinese and both Javanese and Sundanese, compared to GPT-2 tokenization. These results collectively demonstrate that the syllable-based approach is more effective in capturing phonological and morphological pattern alignments that are both historical and areal in nature.

\section{Concluding Remarks \& Further Works}
The unique morphemic and phonetic constructions of Austronesian languages, compared to English which serves as the foundation for the general development of large language models in Generative Artificial Intelligence, have underscored the importance of agglutinative-based tokenization (ToBA-LM). This tokenization approach seeks to follow the "logic" of word formation based on syllabic patterns, which constitute a distinctive characteristic of natural Austronesian languages. This motivates adaptive efforts in conventional tokenization techniques for building language models whose linguistic aspects are syllable-based, as is the general nature of natural Austronesian languages. Case studies for model testing were conducted on corpora from various ethnic languages in the Indonesian Archipelago region. The tokenization approach was carried out with in-depth consideration of how words and phrases are formed from the native scripts of these ethnic groups, encompassing cross-linguistic segmentation and syllable compatibility, as well as in-depth analysis of the alignment processes of the token sequences employed.

This study, in turn, demonstrates the unique tokenization patterns in the constructed model, directed toward a bottom-up implementation of computational language models that can robustly accommodate the development of Indonesian ethnic languages, remaining immune to the emergence of words or phrases that tend to be out of vocabulary—characterized by frequent word segmentations that are not accommodated in conventional tokenizer vocabularies. This nonetheless enhances the efficiency of computational language models when implemented as language models for Generative Artificial Intelligence applications in general, which consistently rely upon and are built solely from English-dominated corpora. By examining the Token per Character (TPC) ratio, a markedly different and inverse pattern is demonstrated between syllable-based agglutinative tokenization and GPT-2. GPT-2 tokenization exhibits low TPC values for English and very high TPC values for ethnic languages in the Indonesian Archipelago, indicating distinct computational characteristics between these two language families. Nevertheless, the tokenization pattern optimized for syllable-based agglutinative processing (ToBA-LM) demonstrates low effectiveness for English. This leads to the conclusion of strong differences in the construction of the developed language models, which certainly may have potential efficiency implications for further large language model implementations.

Furthermore, the diversity and similarity aspects of the ethnic languages used in this study can also be demonstrated, including the proximity between Malay and Banjarese and Minangkabau, which have a relative similarity distance from languages such as Javanese, Sundanese, and Batak Toba. This certainly constitutes an important aspect that can be useful in the construction of individual language models, as well as composite language models based on syllable-based agglutinative approaches within the broader family of ethnic languages in the Indonesian Archipelago, including at a larger scale, languages of the Austronesian family—which represents a conjecture for future work. This includes combinations of this study's implementation with aspects of writing from original manuscripts that use unique scripts from hundreds of other languages in Indonesia, as well as other Austronesian families, employing image pattern recognition approaches from digital scans of manuscripts that facilitate textual deepening of meanings in ancient texts possessing high historical and ancient wisdom aspects spanning hundreds to thousands of years ago.

It is hoped that this study serves as an initial step in the development of advanced computational language models, in the implementation of various Large Language Model (LLM) applications grounded in the unique naturalness of language with all the human aspects that gave birth to it. Strong respect and appreciation for the collective intelligence that gave birth to language as a fundamental entity in human communication is, nonetheless, an important matter in the development of Artificial Intelligence technology that remains humane in its culture and civilization.

\section*{Acknowledgement}

The author thank Institut Teknologi Del, fellows in Bandung Fe Institute, and colleagues in Ina-AI Co. for support. All faults remain authors'.

\bibliographystyle{plain}

\end{document}